\documentstyle[twocolumn,prb,aps]{revtex}
\tightenlines

\begin{document}
\draft

\twocolumn\narrowtext

\title{Pressure Tuning of the Charge Density Wave in the
Halogen-Bridged Transition-Metal ($MX$) Solid
Pt$_2$Br$_6$(NH$_3$)$_4$}
\author{G.S. Kanner, J.Tinka Gammel, S.P. Love, S.R. Johnson, B. Scott,
and B.I. Swanson}
\address{Los Alamos National Laboratory, Los Alamos, NM, 87545}
\maketitle

\vbox{\widetext$~~~~~~~~~~~~~~~~~~~~$\vbox{\mediumtext\vskip -.75truecm

\begin{abstract}
We report the pressure dependence up to 95 kbar of Raman active
stretching modes in the quasi-one-dimensional $MX$ chain solid
Pt$_2$Br$_6$(NH$_3$)$_4$. The data indicate that a predicted
pressure-induced insulator-to-metal transition does not occur, but are
consistent with the solid undergoing either a three-dimensional
structural distortion, or a transition from a charge-density wave to
another broken-symmetry ground state. We show that such a transition
can be well-modeled within a Peierls-Hubbard Hamiltonian.
\end{abstract}


}\narrowtext}

The halogen-bridged ($X$=Cl, Br, I) transition metal ($M$=Pt, Pd, Ni)
compounds ($MX$ solids) have recently attracted a great deal of attention
as paradigms of one-dimensional (1-D) systems \cite{ref01}. Unlike
conjugated polymers, the $MX$ materials typically consist of linear
rather than planar chains, and can be formed as single crystals,
therefore mitigating complications in interpreting experimental results
obfuscated by disorder. Furthermore, $MX$ systems exhibit a wide variety
of broken-symmetry ground states ranging from a strongly Peierls
distorted charge-density-wave (CDW) in PtCl, to a weak one in PtI, to a
spin-density-wave (SDW) in NiBr \cite{ref02}; transitions between these
states are, in fact, a focus of this paper. Theoretical interest in
these materials has also been stimulated by their role as 1-D analogs
of high-temperature Cu-O superconductors \cite{ref03}.

An $MX$ complex is typically represented by $[ML_4][MX_2L_4]Y_4$,
where $L$ is a ligand molecule, such as ethylamine, $X$ represents the
bridging halide, and $Y$ is a counterion, such as ClO$_4^-$. It has been
found that the network of ligands and counterions forms a template to
which the $M$ and $X$ atoms along the chain must adjust \cite{ref04,ref05}.
The result is that the choice of $L$ and $Y$ has a dramatic influence (the
``template effect") on the separation between adjacent $M$ atoms, which
appears to determine the Peierls distortion $\rho$ (the ratio of the
short to long $M$$-$$X$ bond lengths) of the CDW. We have observed
\cite{ref05} that both the frequency of the Raman-active chain mode
$\nu_1$, associated with the symmetric stretching of the axial $X$ atoms
around each $M^{4+}$ atom, and the energy (or charge transfer) gap
$E_g$, increase monotonically as $\rho$ decreases (as the CDW strength
increases), demonstrating the utility of simple optical techniques for
probing ground state properties. The tunability of $E_g$ is obviously
a desirable feature in materials used in optical devices, such as
light-emitting diodes \cite{ref06}.

In addition to chemical substitution, pressure ($P$) can be used to
tune $\rho$, with Raman spectroscopy as a means of detecting the
consequent changes in the geometry of an $MX$ chain through observation
of $\nu_1$. An excellent
candidate for high $P$ Raman studies is
Pt$_2$Br$_6$(NH$_3$)$_4$ (abbreviated PtBrn),
\hfill
an $MX$ solid with neutral chains,
\\ \vskip 2.85truecm \noindent
and, hence, a relatively small number of atoms per unit cell
(see Fig.~1, Ref.~7). This particular material was the subject of two
recent theoretical studies based on local density approximation (LDA)
calculations \cite{ref07,ref08}, one of which predicted the onset of
metallization at $P_C$=89 kbar, induced by uniaxial stress along the
chain axis. The calculations suggested that evidence for this
transition would be a continuous decrease of $\nu_1$ to zero as $P$
increased from zero to $P_C$. Although previous experimental studies
indicated that such a pressure-induced metallization is averted in
charged $MX$ chains ({\it i.e.}, PtCl and PtBr with ethylenediamine ligands
and ClO$_4^-$ counterions) \cite{ref09,ref10,ref11,ref12,ref13,ref14}, it
is essential to recognize that the local chemical environments of the
charged and neutral chains are significantly different from each
other. The importance of the ligands and counterions surrounding the
1-D structure has been demonstrated both experimentally and
theoretically \cite{ref04,ref05,ref07}; in particular it was shown
\cite{ref07} that the NH$_3$ and Br ligands are necessary for enabling
the Peierls distortion in PtBrn. Thus, the behavior of a neutral chain
solid under pressure should not necessarily be expected to mimic that
of charged chain system. We have therefore measured the dependence on
high hydrostatic pressure of $\nu_1$, its first overtone $\nu_1^{1o}$,
and an ``off-chain" symmetric stretching mode $\nu_L$ of PtBrn, up to
the theoretical value for $P_C$. The data indicate that an
insulator-to-metal transition for this neutral chain system does not
occur, implying that a realistic model of the pressure dependence must
at least account for either interchain effects or electron-electron
interactions. In fact, when we include the latter in our calculations
of the uniaxial stress dependence of $\nu_1$, the results of which are
presented below, we can qualitatively reproduce the experimental data.

Raman experiments were performed using a Spex 1877D triple spectrometer
coupled with a 298x1152 CCD array, and an Ar$^+$-pumped CW Ti:Sapphire
laser (Spectra Physics 3900). The average power of the excitation beam
(850 nm) was typically 10 mW, focussed to a spot of about 50 $\mu$m
diameter. Pressures up to 95 kbar were obtained with a Merrill Basset
diamond anvil cell filled with cyclohexane; all high $P$ measurements
were performed at 300K. $P$ was determined from the ruby luminescence
\cite{ref15}, assuming a shift of the ruby lines of 0.365 \AA/kbar
\cite{ref16}. Uncertainties in phonon frequency and cell pressure were
0.5 cm$^{-1}$ and 0.3 kbar, respectively.

The neutral chains were formed from a reaction of a slight excess of
PtBr$_2$(NH$_3$)$_2$ in deionized water with one equivalent of
PtBr$_4$(NH$_3$)$_2$ under pressure and high temperature (7-10 psi and
110-120 C). Evaporation of the solution in a 90 C oven yielded bronze
crystals. The precursors were synthesized from PtCl$_2$(NH$_3$)$_2$
(Aldrich) via a substitution reaction with two equivalents of AgNO$_3$,
forming Pt(NH$_3$)$_2$(NO$_3$)$_2$, and then resubstituting with two
equivalents KBr, forming PtBr$_2$(NH$_3$)$_2$, which can be oxidized to
PtBr$_4$(NH$_3$)$_2$ with Br$_2$.

The Raman spectra at 1 bar and at higher pressures up to 95 kbar (all
at 300K) are shown in Fig.~1. The observed line broadening with $P$ has
contributions from both anharmonicity, and nonhydrostatic conditions in
the cyclohexane above 40 kbar; the latter, however, should not effect
the measured shifts of vibrational frequencies with $P$. We focus on
the $P$ dependence of $\nu_1$, its first overtone $\nu_1^{1o}$, and the
symmetric stretching mode $\nu_L$ of the equatorial Br atoms
perpendicular to the chain. From 1 bar to 30 kbar $\nu_1$ decreases,
in agreement with our expectations for a CDW and the recent
calculations \cite{ref08}. However, between 30 and 40 kbar the slope
is close to zero, and at higher pressures changes sign, such that above
80 kbar $\nu_1$ exceeds its ambient pressure value. This is
qualitatively similar to what was observed for $\nu_1$ under pressure
in charged PtBr and PtCl chains
\cite{ref09,ref10,ref11,ref12,ref13,ref14}. The pressure dependence of
$\nu_1$ is summarized in Fig.~2. We observe that the slope $d\nu_1/dP$
below 30 kbar is much smaller than what is predicted from the LDA
calculations. For instance, at low $P$ the data can be fit with a
linear slope of $-$0.46 cm$^{-1}$/kbar, which extrapolates to zero at 359
kbar.

The initial soft mode behavior of $\nu_1$ with $P$ is consistent with
the template effect \cite{ref04}, and with the recent calculations
\cite{ref08}. As the chain is squeezed, leading to a smaller Pt$-$Pt
separation, and a preferential compression of the long, weaker
Pt$^{2+}$$-$Br bonds, the curvature of the potential well around each Br
atom decreases, with a corresponding decrease of $\nu_1$. At $P$=$P_C$
each Br atom is forced to be at the midpoint between the Pt
atoms, which now become equivalent to each other (Pt$^{3+}$), the solid
is predicted to be a 1-D metal, and the formerly symmetric chain mode
vanishes from the Raman spectrum because it is now antisymmetric with
respect to inversion symmetry in the unit cell, which has been cut in
half. Consequently, this process can be described as a second order
phase transition with $\nu_1$ as the order parameter. From Landau's
theory \cite{ref17}, it follows that,
\vskip-.6truecm
\begin{equation}
\nu_1=A|P-P_C|^{1/2},
\end{equation}
\vskip-.2truecm \noindent
where $A$ is a constant. This $P$ dependence is actually followed by
some materials exhibiting paraelectric to ferroelectric phase
transitions \cite{ref18}. The data from 1 bar to 30 kbar can
also be fit quite well to Eq.~1, yielding $P_C$=196 kbar, even
though such a fit is strictly valid only near $P$=$P_C$.

The most puzzling feature of Fig.~2 is the change in slope above 30
kbar. One possible origin is increased hard-core repulsion between the
Pt and axial Br atoms as they are forced to move toward each other,
causing either alternate metals or halogens to buckle out of the chain,
and leading to a planar, rather than linear structure. Additionally,
we consider the strengthening of interactions perpendicular to the
chains upon compression, an effect not accounted for in Ref.~8. For
example, a zig-zag structure of the chain may also be induced by the
compression of adjacent equatorial Br ligands, which, at ambient
pressure, are already separated by only 3.306\AA, less than twice their
Van der Waals radius (2$\times$1.95\AA=3.90\AA) \cite{ref19}. Instead of
moving closer together, these atoms may move parallel to the chain, in
opposite directions, pulling the axial Br atoms off the chain, and,
perhaps, even stretching their bonds to the Pt atoms. We therefore
conjecture that either of the above two effects could cause $\nu_1$ to
increase with $P$ as optical phonons typically do in 3-D solids in the
absence of phase transitions \cite{ref16}. The importance of 3-D
interactions was also revealed by LDA calculations based on PtCl
\cite{ref08}, which simulated a reduction in interchain distance for a
fixed Pt$-$Pt separation, and suggested that an increase in overlap
between the orbitals of the off-axis ligands and those of the axial Pt
and Cl atoms, would increase $E_g$. We then expect from the template
effect \cite{ref04}, that this would be associated with an increase in
$\nu_1$.

The $P$ dependence of $\nu_1$ in PtBrn is not typical of all $MX$ solids,
however. In PtI, for example, a system with a very weak CDW, $\nu_1$
increases with $P$ from 1 bar to 30 kbar \cite{ref10}, opposite to the
low $P$ dependence of $\nu_1$ in PtBrn. Nevertheless, the hardening of
$\nu_1$ with $P$ that occurs in PtBrn and charged PtBr and PtCl at high
$P$, when the Peierls distortion is ostensibly diminished, may be
consistent with the low $P$ behavior of PtI, which is weakly distorted
at only 1 bar.

The appearance of two pressure-induced modes at 177 and 192 cm$^{-1}$
($P$=53 kbar, Fig.~1) is suggestive of a change in crystal symmetry
with $P$, yet the threshold pressure for these modes was sample
dependent, occurring as low as 17 kbar. These modes therefore do not
appear to be correlated with the transition at 35$\pm$5 kbar, which showed
little dependence on sample. However, modes were observed in charged
PtBr at 174 and 182 cm$^{-1}$ with 2.41 eV excitation, and attributed
to polaronic defects, the Raman intensities of which were presumably
enhanced by optical transitions from the upper halide ($X$) band to a gap
state between the metal ($M$) bands \cite{ref20}. In the mixed-halide
compound PtCl$_x$Br$_{1-x}$ ($x$=0.75), modes were also observed upon
photolysis at 2.54 eV at 171 and 181 cm$^{-1}$, and ascribed to the
symmetric stretching of the axial Br atoms around Pt$^{4+}$ on short
chain segments \cite{ref21}. The observation of new modes in this
study only at the highest pressures, for fixed excitation energy (1.46
eV), is suggestive of redshifts of the interband transition energies
with pressure, such as the decrease of $E_g$ with $P$ in a PtCl
compound \cite{ref11}. Thus, for example, resonance with the defect
modes may occur only when pressure reduces the separation in energy
between the upper $X$ band and the lower $M$ band such that the $X$ to
polaron level transition energy approaches $E_{ex}$. Alternatively,
pressure may lead to a reduction of $E_g$ of the short chain segments,
which have optical gaps greater than that of the ``infinite chain"
\cite{ref22}; the symmetric stretching mode for these segments is then
enhanced when $E_g\approx E_{ex}$.

Although the structural distortions mentioned above are possible
impediments to metallization, we must also consider that pressure
forces the CDW to undergo a transition to a different broken-symmetry
ground state. We have tried to account for this effect by calculating
the band structure for PtBrn within a 2-band, 3/4-filled tight-binding
Peierls-Hubbard (PH) model \cite{ref23} with linear electron-phonon
couplings and springs using the Hamiltonian,
\vskip-.6truecm
\begin{eqnarray}
 H &=&
\sum_{i,\sigma} \bigg[ (-t_0+\alpha \delta_i) (c^{\dag}_{i,\sigma}
 c_{i+1,\sigma} + c^{\dag}_{i+1,\sigma} c_{i,\sigma})
\nonumber \\ & &
{}~~~~~~~~~~~~~~
+ (e_i-\beta_i(\delta_i+\delta_{i-1})) n_{i,\sigma} ~~~~~~~ \bigg]
\nonumber \\ & &
 + \sum_i U_i n_{i,\uparrow} n_{i,\downarrow}
\nonumber \\ & &
+ \sum_{i} \bigg[ \frac {1}{2}M_i \dot{v}^2_i
 + \frac {1}{2}K_{MX} (\delta_i-{\rm a}_0)^2
\nonumber \\ & &
{}~~~~~~~~~~~~~~
+\frac {1}{2}\kappa_i (\delta_i+\delta_{i-1}-2{\rm a}_0)^2
\nonumber \\ & &
{}~~~~~~~~~~~~~~
+ PA_{\perp} (\delta_i-{\rm a}) ~~~~~~~~~~~~~~ \bigg]
\end{eqnarray}
\vskip-.2truecm \noindent
with nearest-neighbor hopping $t_0$, on-site energy ($e_i$,
$e_M$=$-e_X$=$e_0$$>$0), on-site ($\beta$) and intersite
($\alpha$) electron-phonon coupling to the deviation $\delta_i$ at site
$i$ from the reference ($P$=0) lattice constant ${\rm a}_0$, an on-site ($U$)
Hubbard term, effective near-neighbor ($K$) and next-nearest-neighbor
($\kappa$) springs (to model interactions not explicitly included), and
pressure $P$. $\beta$ may also effectively be viewed as modeling the
distance dependence of the Coulomb interaction \cite{ref24}. The
lattice distortion is determined self-consistently. This implies that
with pressure, $\delta_i$ acquires an $i$-independent component,
$d_0(P)=\sum_i \delta_i/N<0$, corresponding to a reduction in the
lattice constant, ${\rm a}^{eff}(P)={\rm a}_0+d_0(P)$, and an increase in the
effective hopping integral, $t_0^{eff}(P)=t_0-\alpha d_0(P)$.

Thus far, we have studied only uniaxial pressure (fixed perpendicular
cross-sectional area $A_\perp$), and ignored interchain interactions,
which could give rise to a neutral-ionic transition similar to the case
in mixed stack charge-transfer materials \cite{ref14}. For large
deviations from the ground state geometry, nonlinear electron-phonon
couplings and springs should also be taken into account. The agreement
is, however, surprisingly good even without such corrections
\cite{ref25}. Our modeling indicates that electron-electron
correlations and interchain interactions are relatively weak in PtBrn,
and thus treatment of the single chain Peierls-Hubbard Hamiltonian
within mean field approximation should be valid for PtBrn parameters,
though more accurate treatment may be necessary in the high-$P$ regime,
in which the lattice distortion may vanish, and interchain interactions
become increasingly important.

Using the $P$=0 LDA results \cite{ref07} for the band structure and our
experimental results, we have determined a parameter set for Eq.~2
that correctly reproduces the low $P$ dependence of $\nu_1$, as shown
in Fig.~2. With these parameters we predict an initial reduction in
the amplitude of the CDW lattice distortion, followed by a transition
to a pure SDW phase \cite{ref26} near 40 kbar that gives rise to a
turnaround in $d\nu_1/dP$, in agreement with our observations, but in
contrast to LDA calculations for $P$$>$0 \cite{ref08}. However, we expect, for
negligible spin-phonon interactions, that $\nu_1$ would become Raman
inactive in the SDW phase. We are therefore investigating other
parameter sets that lead to a transition to a commensurate spin-Peierls
phase \cite{ref27}, in which $\nu_1$ would remain symmetry allowed.

We find that the first overtone of $\nu_1$ provides further evidence of
a pressure-induced phase transition. In Fig.~3 we plot the deviation
(from harmonicity), defined as $\Delta\nu={1\over2} \nu_1^{1o}-\nu_1$,
as a function of $P$. $\Delta\nu$ is positive for
almost all $P$, and exhibits a peak at 35 kbar, which is in the region
where $\nu_1$ {\it vs.} $P$ is flat. For isolated molecules, overtones
are ordinarily at lower frequencies than integral multiples of the
fundamental due to anharmonicity \cite{ref28}. However, for solids, the
first overtone has only to satisfy the sum rule
$\vec q_1 + \vec q_2 \approx 0$,
where $q_i$ is the wavevector of each emitted phonon. This
implies that the overtone can have contributions from phonons
throughout the Brillouin zone, and not just at $q$=0. Since the
dispersion curve on which $\nu_1$ lies slopes upwards from $q$=0 to
$q=\pi/{\rm a}_0$ \cite{ref29}, it is plausible that the high frequency of
the overtone is due to phonons at $q$$>$0. The most likely candidates
have wavevectors corresponding to a high density of states $g(q)$.
Since $g(q)\sim|\partial\omega/\partial q|^{-1}$ in 1-D, where
$\omega$ is the angular phonon frequency, the greatest contributions
would correspond to extrema of the dispersion curve, such as at $q=0$
and $q=\pi/{\rm a}_0$ at ambient pressure. However, as $P$ increases, a kink
may develop in the phonon dispersion curve between zone center and zone
boundary, giving a high density of states in that region. Such an
anomaly was observed via neutron scattering for a soft phonon mode in
SrTiO$_3$ as a function of temperature \cite{ref30}. In our case, the
kink would probably be correlated with either a change in crystal
structure, or a transition from the CDW to another broken symmetry
ground state.

We have also investigated the $P$ dependence of another mode $\nu_L$,
which has a frequency at 1 bar of 204 cm$^{-1}$, close to that
(207 cm$^{-1}$) of the Pt$-$Br vibration of the isolated (Pt$^{2+}$) molecule
\cite{ref31}, and, relative to $\nu_1$, is enhanced by light polarized
perpendicular to the chains. We therefore attribute $\nu_L$ to the
symmetric stretch of the equatorial Br atoms around each Pt. As Fig.~2
shows, $\nu_L$ {\it increases} linearly (0.37 cm$^{-1}$/kbar) up to
95 kbar; this positive slope is in stark contrast to the softening of the
chain mode $\nu_1$ at low pressures, and suggests that a CDW does not
exist parallel to the equatorial Br ligands. Initially, we had hoped
that compression would lead to delocalization along the axis defined by
the Pt$-$equatorial Br atom bonds, and that a CDW would form as a
result. This would be facilitated by the close proximity of adjacent
equatorial Br atoms to each other, but would require chain-to-chain
ordering, manifested as an alternation in the Pt valences. However,
the increase in $\nu_L$ with pressure is consistent with either a lack
of ordering, or, as in our interpretation of the high pressure
dependence of $\nu_1$, a 3-D distortion of the equatorial Br ligands
occurs as we attempt to press them together.

In conclusion, we have shown that the pressure dependence of the Raman
active chain mode in Pt$_2$Br$_6$(NH$_3$)$_4$ at low $P$ is in
qualitative agreement with the template effect, and with the
theoretical predictions for the quenching of the CDW.  At higher $P$,
however, metallization is obstructed by what may be a 3-D distortion of
the chains, or a transition to another broken-symmetry state such as a
spin-Peierls phase, driven by electron-electron interactions.
We believe the present results, in conjunction with those for charged
chains, are suggestive of a basic universality in the
high pressure response of MX solids.  Identification of the origin of this
phenomenon may require high pressure X-ray diffraction and
magnetic measurements on these materials.

We would like to thank M. Alouani, A.R. Bishop, R. Donohoe, A. Saxena
and D. Schiferl, for useful discussions. This work was supported by
the Division of Materials Science of the Office of Basic Energy
Sciences of the DOE.


\vskip -.5truecm
\noindent\underbar{\hbox to  \columnwidth {\hfil}}

\begin{figure}\caption{\label{fig1}
Raman spectra excited at 850 nm at various pressures
from 1 bar to 95 kbar at 300 K.
}\end{figure}
\begin{figure}\caption{\label{fig2}
Pressure dependence at 300K of the Raman-active chain mode $\nu_1$
(solid circles), and $\nu_L$ (hollow circles), the symmetric stretching
mode of the equatorial Br ligands. The solid line through $\nu_L$ is a
linear fit (see text). The solid line through $\nu_1$ is a theoretical
fit based on Eq.~2 using the parameters $t_0$=0.98 eV, $e_0$=0.69 eV,
$\alpha$=1.56 eV/\AA, $\beta_M$=1.37 eV/\AA, $\beta_X$=$-$4.10 eV/\AA,
$U_M$=1.60 eV, $U_X$=0.32 eV, $K_{MX}$=6.14 eV/\AA$^2$ and
$K_{MM}$=0.61 eV/\AA$^2$.
}\end{figure}
\begin{figure}\caption{\label{fig3}
Pressure dependence of $\Delta\nu$. The solid horizontal line
represents $\Delta\nu$ = 0, corresponding to a purely harmonic crystal
with no density of states effect.
}\end{figure}


\begin{references}
\bibitem{ref26}
When 3-D couplings and corrections to mean field are added, whether this
``SDW" phase is antiferromagnetic, an actual SDW, or an
antiferromagnetic plus CDW spin-Peierls phase (distinct from the ``MXP"
spin-Peierls phase -- both types of spin-Peierls phases can occur --
see Ref.~27) will depend on both chain lengths and the strength of the
3-D couplings. However, such considerations should not change the
qualitative arguments here.
\end{references}
\end{document}